\documentstyle[aps,prl,multicol,epsfig]{revtex} 
\begin{document}
\title{ On gauge-invariant Green function in $ 2+1 $ dimensional QED }
\author{ Jinwu Ye }
\address{ Department of Physics and Material Research Institute, The Pennsylvania State University, University Park,
   PA, 16802 }
\date{\today}
\maketitle
\begin{abstract}
     Both the gauge-invariant fermion Green function and gauge-dependent conventional Green function
  in $ 2+1 $ dimensional QED are studied in the large $ N $ limit.
  In temporal gauge, the infra-red divergence of gauge-dependent Green function is found to be regulariable,
  the anomalous dimension is found to be $ \eta= \frac{64}{ 3 \pi^{2} N} $.
  This anomalous dimension
  was argued to be the same as that of gauge-invariant Green function.
  However, in Coulomb gauge, the infra-red divergence of the gauge-dependent Green function
  is found to be un-regulariable, anomalous dimension
  is even not defined, but the infra-red divergence is shown to be cancelled in any gauge-invariant
  physical quantities. The gauge-invariant Green function is also studied directly in 
  Lorentz covariant gauge and the anomalous dimension is found to be the same  as that
  calculated in temporal gauge.
\end{abstract}
\begin{multicols}{2}
\section{ Introduction}

    In high energy physics, $ 2+1 $ dimensional massless Quantum Electro-Dynamics ( QED3 ) 
   was studied as an interesting theoretical toy model to gain
  insight into dynamical symmetry breaking of $ 3+1 $
  dimension QED \cite{qed}. In condensed matter systems, QED3 may appear as an effective
  low energy theory describing doped Mott insulators \cite{flux,brad,u1}.
  The connection of dynamical mass generation in QED3 with Anti-ferromagnets was also discussed in
  \cite{brad}. QED3 with possible
  Chern-Simon term may also describe Fractional Quantum Hall (FQH) transitions \cite{frad,hlr,chen,subir,disorder}. 

   In the two component notations suitable for describing FQH transitions \cite{hlr,chen,subir,disorder},
  the standard $ 2+1 $ dimensional massless Quantum Electro-Dynamics ( QED3 )  Lagrangian in Euclidean space is:
\begin{equation}
   {\cal L} =  \bar{\psi}_{a} \gamma_{\mu} (\partial_{\mu}
    -i e  a_{\mu} ) \psi_{a} + \frac{1}{4 } ( f_{\mu \nu} )^{2}
\label{qed}
\end{equation}
  Where the Dirac matrices in three dimensions are $ \gamma_{\mu}= i \sigma_{\mu} $ with $ \sigma_{\mu} ( \mu=1,2,3 ) $
 being three Pauli matrices \cite{chen} and $ a=1,\cdots, N $ are  $ N $ species of Dirac fermion.

   The conventional Green function in QED3 is defined as:
\begin{equation}
  G(x_{1}, x_{2}) = < \psi(x_{1}) \bar{\psi}(x_{2} )>   
\label{non}
\end{equation}

  In momentum and real space, the fully-interacting Green function in Eqn.\ref{non}
  takes the form:
\begin{equation}
  G( k )= \frac{ i k_{\mu} \gamma_{\mu} }{ k^{2 + \eta} },~~~~~~~ G( x )= \frac{ \gamma_{\mu} x_{\mu} }{ x^{3-\eta} }
\end{equation}
  where the convention we chose for the sign of the anomalous dimension $ \eta $ is that
  a positive $ \eta $ slows down the algebraic decay of the Green function in real space,
  $ \eta $ can be calculated by standard  Renormalization Group (RG) by
  extracting UV divergences. Unfortunately, this Green function is not gauge invariant, $ \eta $
  depends on the fixed gauge in which the calculation is done.

  The gauge invariant Schwinger Green function which is the focus of this paper is
\begin{equation}
  G^{inv}(x_{1}, x_{2} ) =
   < \psi(x_{1}) e^{i e \int^{x_{2}}_{x_{1}} a_{\mu}(\xi) d \xi_{\mu} } \bar{\psi}( x_{2} )>
\label{inv}   
\end{equation}
   where the inserted Dirac string makes the Schwinger Green function gauge invariant.

  $ G^{inv} $ depends on the the integral path  $ \cal{ C} $ from $ x_{1} $ to $ x_{2} $. For simplicity
  reason, we take $ {\cal C} $ to be a straight line. 

   As far as I know, the importance of gauge-invariant Green function in condensed matter
  system was first realized in FQH system \cite{he}. By a singular gauge transformation which attaches
  two flux quantums to each electron,
  an electron in an external magnetic field was mapped to a composite fermion in a
  reduced magnetic field. Although transport properties which are directly related to
  two particle Green functions can be directly studied in the composite fermion language,
  the tunneling density of states which is directly related to
  the single particle electron Green function is much more difficult to study.
  In fact, the single particle electron Green function
  is equal to the gauge-invariant Green function of the composite fermion 
  which was evaluated for non-relativistic fermions by phenomelogical arguments in Ref.\cite{he}.
 
    Most recently, the importance of gauge-invariant Green function of fermion 
    to Angle Resolved Photo-Emission  (ARPES) data in high temperature superconductors
    was discovered independently by Rantner and Wen (RW) \cite{wen} and the author \cite{thermal}
    in different contexts.
    Starting from $ SU(2) $ gauge theory of doped Mott insulators
    \cite{u1}, RW discussed the relevance of this gauge-invariant Green function to ARPES data.
    They also pointed out that in temporal gauge, the equal-space gauge-invariant Green function in Eqn.\ref{inv}
    is equal to that of the conventional gauge dependent one in Eqn.\ref{non}.
    Furthermore, they calculated the anomalous
    dimension of the conventional Green function in temporal gauge by large $ N $ expansion. However,
    both the sign and magnitude of their result differs from that achieved in this paper.
    Starting from a complementary (or dual ) approach pioneered  by Balents {\sl et al} \cite{balents},
    the author studied  how  quantum \cite{quantum} or thermal \cite{thermal} fluctuations generated
    $ hc/2e $ vortices
    can destroy $ d $-wave superconductivity and evolve the system into underdoped regime at $ T=0 $
     or pseudo-gap regime at finite $ T $. 
    The author found that only in weakly type II $ d $-wave superconductors where the interaction
    between vortices is short-ranged, the effective low energy
    theory is described by QED3. For underdoped cuprates which are between quantum anti-ferromagnet and
    strongly type II $ d $ wave superconductors, the correct $ T=0 $ theory is still unknown. However,
   in the vortex plasma regime around the {\em finite } temperature Kosterlize-Thouless transition \cite{emery}, the vortices
   can be treated by classical hydrodynamics. By Anderson singular gauge transformation which attaches
   the flux from the classical vortex to the quasi-particles of $ d $-wave superconductors 
   \cite{and,static,russ,analogy}, the quasi-particles (spinons) are found to move in a random 
   magnetic field generated by the classical vortex plasma \cite{thermal}.
   The electron spectral function $ G(\vec{x}, t) =<C_{\alpha}(0,0) C^{\dagger}_{\alpha}(\vec{x}, t)> $
   is the product of the classical vortex correlation function and
   the {\em gauge invariant} Green function of the spinon in the random magnetic field.
   Obviously, this static gauge invariant Green function is different from that calculated in this
   paper and will be investigated by completely different methods in separate publication \cite{bose}.

    In this paper, for {\em pure} theoretical interests, but with {\em no} application to
  high temperature superconductors in mind, we study the gauge-invariant fermion Green functions
  Eqn.\ref{inv} in large $ N $ limit.
  By applying the methods developed to study clean \cite{subir} and disordered \cite{disorder}
  FQH transitions and superconductors to insulator
  transitions \cite{si},
  we study the gauge-dependent Green function both in temporal gauge and in Coulomb gauge.
  In both gauges, we regularize the UV divergences in two different cut-offs.
  In both gauges, we also run into expected IR singularities. 
  We also make detailed comparisons between the natures of the IR singularities in these two gauges.
  In temporal gauge, the infrared divergence is in the middle of the contour integral along the real axis.
   From physical prescription, it can be regularized by deforming the contour into complex plane,
  the anomalous dimension is found to be $ \eta= \frac{64}{ 3 \pi^{2} N} $.
  From RW's observation that equal-space
  gauge invariant Green function is the same as the equal space gauge dependent one in temporal gauge,
  we expect this anomalous dimension to be the same as that of gauge-invariant Green function.
  However, in Coulomb gauge, the infra-red divergence is at the two ends of the contour integral
  along the real axis, therefore un-regulariable, anomalous dimension
  is even not defined. This should not cause too much concern, because the anomalous dimension
  in Coulomb gauge does not corresponds to any physical gauge-invariant quantity.
  This infra-red divergence can be shown to be canceled in any physical gauge invariant quantities such as
  $ \beta $ function and correlation length exponent $ \nu $ as observed in Refs.\cite{subir}.
  We also study the gauge-invariant Green function directly by Lorentz covariant
  calculation with different gauge-fixing parameters and find that the exponent  is independent
  of the gauge fixing parameters and is exactly the same as that found in temporal gauge.
   
   The paper is organized as follows. In the next section, we briefly review the previous work
  on gauge dependent Green function  in
  Lorentz covariant Landau gauge \cite{qed}. In section III,  we calculate the gauge dependent Green function
  in temporal gauge with two different cut-offs. In section IV, 
  we discuss the gauge dependent Green function in Coulomb gauge also with two different cut-offs.
  We make detailed comparisons with the calculations done in temporal gauge.
  In section V, we calculate directly the gauge invariant Green function in Lorentz covariant gauge
  with different gauge fixing parameters. Finally, we reach conclusions.

\section{ The calculation in Landau gauge }

   In this section, we will briefly review the previous work on the gauge-dependent Green Eqn.\ref{non} in
  QED3. There is nothing new in this section.
  But it is  constructive to compare the calculations done in  Lorentz covariant gauge in this section
  with those done in temporal gauge in section III and in Coulomb gauge in section IV. It is also a useful
  first step for the Lorentz covariant calculation of gauge-invariant Green function Eqn.\ref{inv} to be
  presented in Sec. V.

  Integrating out $ N $ pieces of fermions generates an additional dynamic quadratic term
  for the gauge field:
\begin{equation}
  {\cal S }_{2} = \frac{1}{2} \int \frac{d^{3} k}{ (2 \pi)^{3} } a_{\mu}(-k) \Pi_{\mu \nu}( k ) a_{\nu}( k )
\label{gen}
\end{equation}

   To one-loop order:
\begin{equation}
   \Pi_{\mu \nu}= \frac{N}{16} k ( \delta_{\mu \nu} -\frac{k_{\mu} k_{\nu}}{k^{2}} )
\end{equation}

   Adding the non-local gauge fixing term and combining
  the original Maxwell term in Eqn.\ref{qed} with the generated dynamic gauge field terms in Eq, \ref{gen}, the
  propagator of the gauge field is \cite{qed}:
 
\begin{equation}
   D_{\mu \nu} = \frac{ 1 }{ k^{2} + \frac{N e^{2} }{16} k}
   ( \delta_{\mu \nu} - (1-\alpha) \frac{k_{\mu} k_{\nu}}{k^{2}} )
\end{equation}

   In general, the Ultra-Violet ( UV ) $ k \rightarrow \infty $ behavior of this propagator is controlled by the
   original Maxwell term $ \sim 1/ k^{2} $, the Infra-Red (IR) one $ k \rightarrow 0 $ is dictated by
   the generated term $ \sim 1/k $.
   In this paper,  we consider large $ N $ limit, such that $ N e^{2} \gg k $, so we neglect the original
   Maxwell term. The above propagator becomes:
\begin{equation}
   D_{\mu \nu} = \frac{ 16 }{ N e^{2} } \frac{1}{k}
   ( \delta_{\mu \nu} - (1-\alpha) \frac{k_{\mu} k_{\nu}}{k^{2}} )
\label{alpha}
\end{equation}

   In Landau gauge $ \alpha=0 $, the one-loop fermion self-energy Feymann diagram
   of the Green function Eqn.\ref{non} is given by:

\vspace{-1.5cm}

\epsfig{file=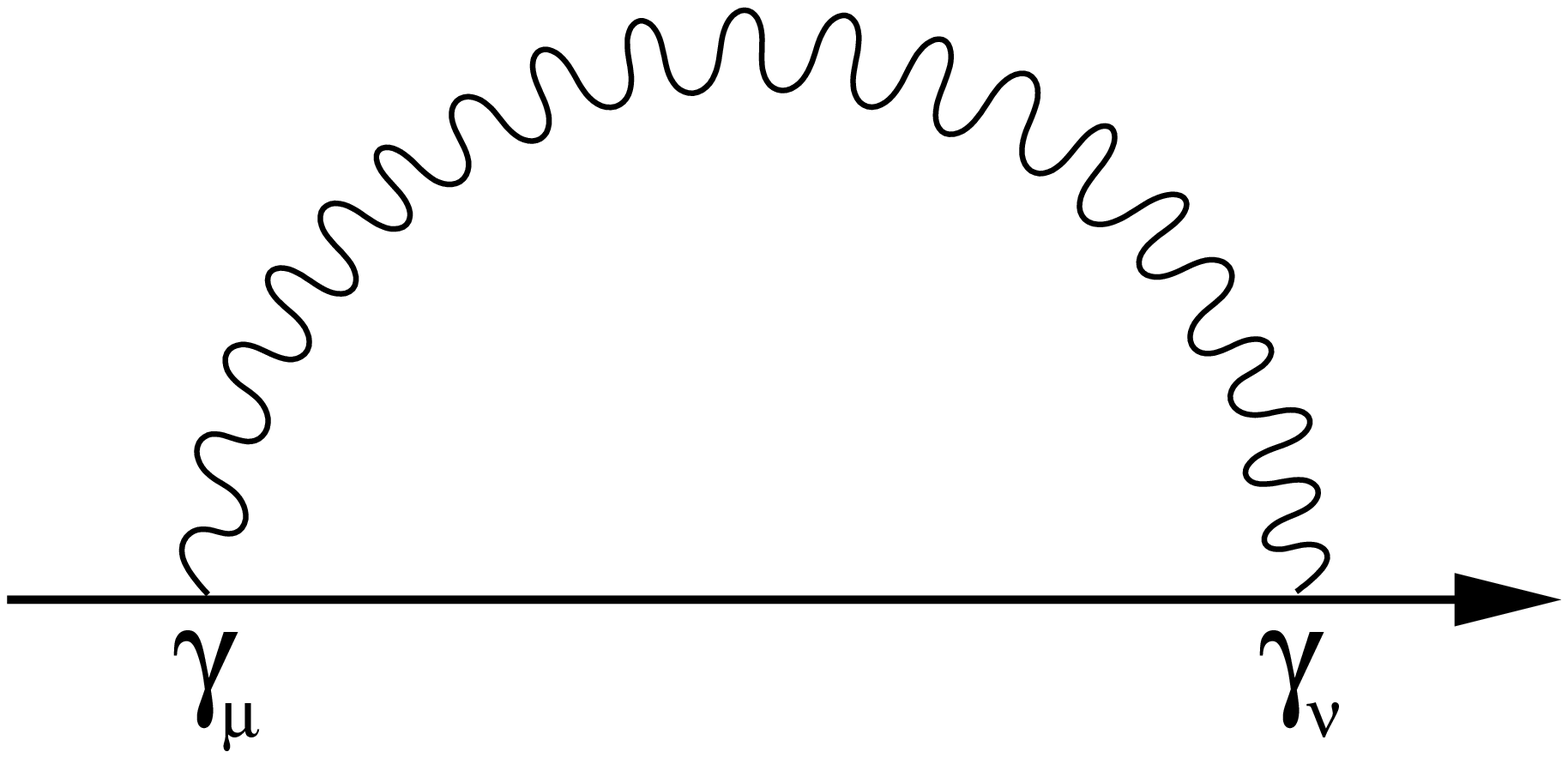,width=2.5in,height=2.5in,angle=0}

\vspace{-2.0cm}

{\footnotesize {\bf Fig 1:} The fermion self-energy diagram in Landau gauge }

\vspace{0.25cm}

   The corresponding expression is
\begin{equation}
   \Sigma (k)= -i \frac{16}{N} \int \frac{ d^{3} q }{ (2 \pi)^{3} }
     \gamma_{\mu} \frac{ \gamma_{\lambda} (k-q)_{\lambda} }{ (k-q)^{2} } \gamma_{\nu}
    \frac{1}{q} ( \delta_{\mu \nu} - \frac{q_{\mu} q_{\nu}}{q^{2}} )
\end{equation}

   By extracting UV divergence,
  we find the anomalous exponent $ \eta= \frac{16}{N} \frac{1}{ 3 \pi^{2} } $.
  If we choose different gauge fixing parameter $ \alpha $, we would get $ \eta = \frac{16}{ N}
  \frac{1}{ 3 \pi^{2} }-\frac{8 \alpha}{ N \pi^{2} } $.
  $ \eta $ is clearly gauge-dependent quantity and its physical meaning in any Lorentz covariant
  gauges which includes Landau gauge is not evident. In the following section, we will calculate
  $ \eta $ in two Lorentz non-covariant gauges: temporal gauge and Coulomb gauge.

\section{ The calculation in Temporal gauge}

     As stated in the introduction, the main focus of this paper is the gauge-invariant Green function Eqn.\ref{inv}.
  In temporal gauge, the equal-space
  gauge invariant Green function is the same as the gauge dependent one \cite{wen}.
  This can be seen easily as follows.

   At equal space, Eqn.\ref{inv} becomes: 
\begin{equation}
  G^{inv}(0, \tau) =
   < \psi( \vec{x}, 0 ) e^{i e \int^{ \tau }_{0} a_{0}(\xi) d \xi_{0} } \bar{\psi}(\vec{x}, \tau )>   
\end{equation}
 
    In temporal gauge $ a_{0} =0 $ \cite{cosmology}, the Dirac string drops out, the equal-space gauge-invariant
  Green function coincides with the equal-space conventional one:
\begin{equation}
  G^{inv}(0, \tau) =
   < \psi( \vec{x}, 0 ) \bar{\psi}(\vec{x}, \tau )> = G( 0, \tau)    
\end{equation}

   The strategy is to calculate the conventional Green function in temporal gauge $ a_{0}=0 $ and then see what
   we can say about the gauge invariant Green function.
   As shown in the last section, in Lorentz covariant gauge Eqn.\ref{alpha}, there is no IR divergences to worry about,
   we need only to regularize the UV
   divergence. However, in temporal gauge and Coulomb gauge which break Lorentz invariance, 
   we run into both UV and IR divergences as expected.
   Care is needed to regularize these plaguey IR divergences in a physical way.

  With the notation $ K_{\mu}= ( \vec{k}, k_{0} ) $ ( namely $ k_{0} $ is along $ z $ direction ),
  in $ a_{0} =0 $ gauge, we can invert Eqn.\ref{gen} to find the propagator:
\begin{equation}
   D_{i j}(K) = \frac{ 16 }{ N e^{2} } \frac{1}{K} ( \delta_{i j} + \frac{k_{i} k_{j}}{k_{0}^{2}} )
\end{equation}
     where $ K^{2}= k^{2}_{0} + ( \vec{k} )^{2}  $.

   The one-loop fermion self-energy Feymann diagram is

\vspace{-1.5cm}

\epsfig{file=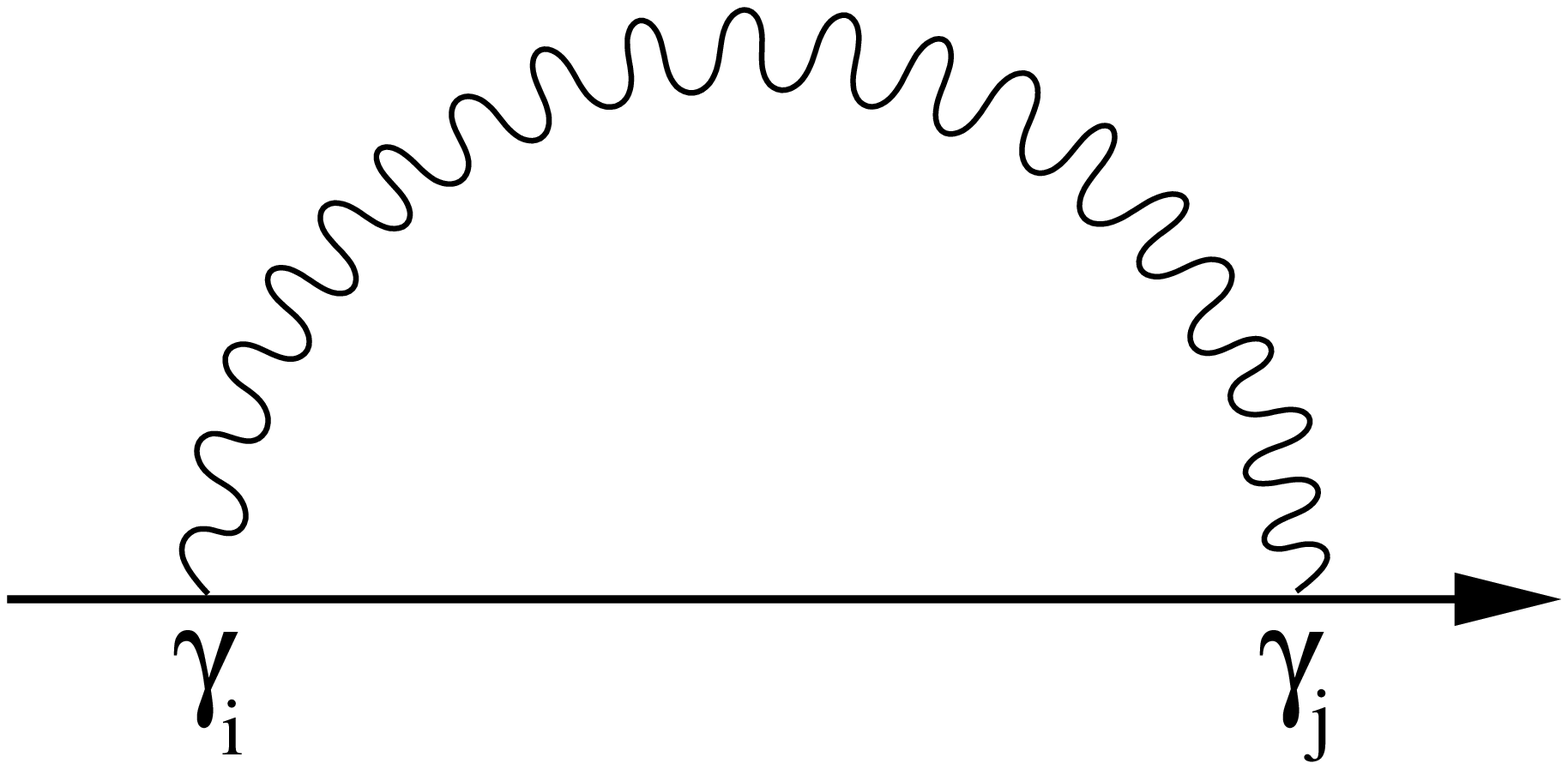,width=2.5in,height=2.5in,angle=0}

\vspace{-2.0cm}

{\footnotesize {\bf Fig 2:} The fermion self-energy diagram in temporal gauge }

\vspace{0.25cm}

    The corresponding expression is
\begin{equation}
   \Sigma (K)= -i \frac{16}{N} \int \frac{ d^{3} Q }{ (2 \pi)^{3} }
     \gamma_{i} \frac{ \gamma_{\mu} (K-Q)_{\mu} }{ (K-Q)^{2} } \gamma_{j}
    \frac{1}{Q} ( \delta_{i j} + \frac{q_{i} q_{j}}{q_{0}^{2}} )
\end{equation}

    By using standard $ \gamma $ matrices algebra and suppressing the prefactor $ -i \frac{16}{N} $,
    we can simplify the above equation to:
\begin{eqnarray}
   \Sigma (K) & = & \gamma_{0}  \int \frac{ d^{3} Q }{ (2 \pi)^{3} }
    \frac{ (k-q)_{0} }{ (K-Q)^{2} Q} ( 2 + \frac{\vec{q}^{2}}{ q^{2}_{0} } )
                  \nonumber    \\
      & + & \gamma_{i} \int \frac{ d^{3} Q }{ (2 \pi)^{3} }
     \frac{ \vec{q}^{2} ( k+q)_{i} -2 \vec{q} \cdot \vec{k}
      q_{i}  } { (K-Q)^{2} Q q^{2}_{0} } 
\label{sim}
\end{eqnarray}

     In the following, we will try to evaluate $ \Sigma ( K ) $ by two different cut-offs.

\subsection{ Cut-off in momentum and energy $ Q < \Lambda $}

 We choose the external momentum $ K $ to be along $ z $ axis, namely $ K_{\mu}= (0, 0, 1) $,
 then in spherical coordinate system, the components
 of integral vector $ Q $ are:
 $ q_{0}= Q \cos \theta, q_{1} =Q \sin \theta \cos \phi, q_{2}=Q \sin \theta \sin \phi $, the second term
 in Eqn.\ref{sim} vanishes due to the $ \phi $ integral. The logarithmic divergence of the first term becomes:
\begin{equation}
  \frac{\gamma_{0} k_{0} } { (2 \pi)^{2} } \int^{1}_{-1} d x (1-2 x^{2} )(1 + x^{-2} ) \log \Lambda
\end{equation}
    where $ x= -\cos \theta $.

    The integral can be rewritten as:
\begin{equation}
   \int^{1}_{-1} \frac{ d x}{  x^{2} } -\frac{10}{3}
\label{div}
\end{equation}
  
    As expected, we run into IR divergence at $ x=0 $ \cite{geo} which is in the middle point of
  the contour integral on the real axis from $ -1 $ to $ 1 $. Fortunately, by physical
  prescription, we can avoid the
  IR singularity at $ x=0 $ by deforming the contour as shown in Fig.3 
\vspace{-1.5cm}

\epsfig{file=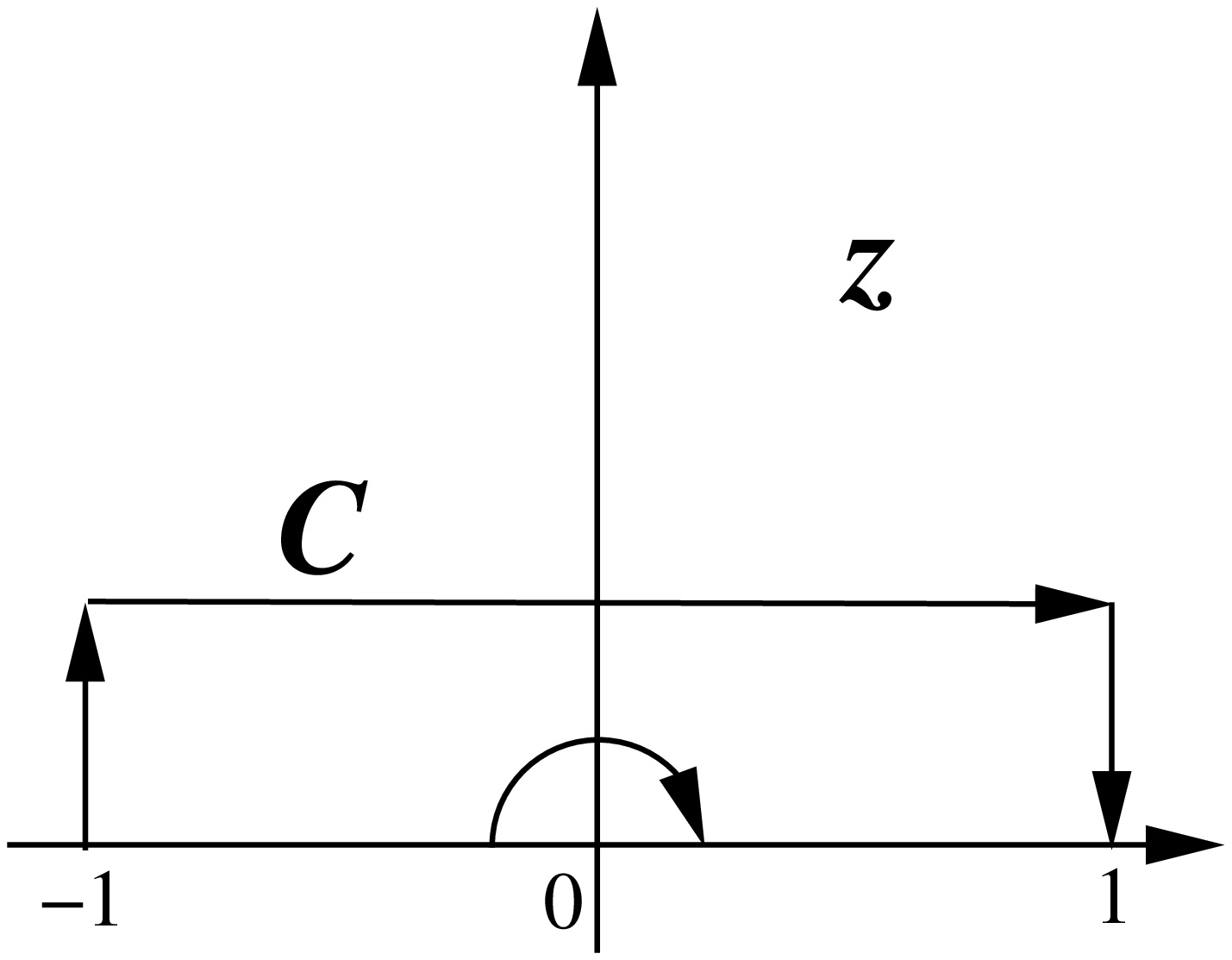,width=3.2in,height=2.5in,angle=0}

\vspace{-2.0cm}

{\footnotesize {\bf Fig 3:} The contour path $ {\cal C } $ to bypass the IR singularity in temporal gauge }

\vspace{0.25cm}

    The divergent part in Eqn.\ref{div} becomes:

\begin{equation}
   \int^{1}_{-1} \frac{ d x }{ x^{2} } = \int_{ C } \frac{ d z} { z^{2} } =
   \int^{1}_{-1} \frac{ d x }{  ( x+ i \epsilon )^{2} } =-2  
\end{equation}

  Putting back the prefactor $ -i \frac{16}{N} $, we get the final answer:
\begin{equation}
   -i \frac{16}{N} \frac{\gamma_{0} k_{0} }{ 4 \pi^{2} } ( -\frac{16}{3} ) \log \Lambda =
   i \gamma_{0} k_{0} \frac{64 }{ 3 \pi^{2} N } \log \Lambda
\label{exp1}
\end{equation}

    We can identify the anomalous dimension as \cite{ramond}
\begin{equation}
  \eta= \frac{64 }{ 3 \pi^{2} N } 
\label{exp2}
\end{equation}

    We expect this is the correct anomalous dimension of the gauge-invariant Green function Eqn.\ref{inv}.
  Note that it has the {\em same } sign as that calculated in Landau gauge.

\subsection{ Cut-off in momentum space $ q < \tilde{\Lambda} $ only}
   
   In last subsection, we choose a cut-off $ \Lambda $ in $ Q $, in this subsection, we introduce
  a alternate cut-off $ \tilde \Lambda $ only in momentum space $ \vec{q} $, but integrate the frequency $ q_{0} $
  freely from $ -\infty $ to $ \infty $. The motivation to choose different cut-off ( or different
  renormalization scheme ) is to check if the exponent Eqn.\ref{exp2} is universal.

  By setting $ \frac{q_{0}}{q} = \frac{2 x}{ 1-x^{2} } $ and
  extracting the logarithmic UV divergences, we find:
\begin{eqnarray}
   \Sigma (k) & = &  \frac{\gamma_{0} k_{0}}{ 4 \pi^{2}}  \int^{1}_{-1}
    dx \frac{ 1-x^{2} }{ (1 +x^{2} )^{4} } \frac{ (1+ x^{4})^{2}-36 x^{4} }{ 2 x^{2} }
     \log \tilde{\Lambda}             \nonumber    \\
    & + & \frac{ \gamma_{i} k_{i} }{ 4 \pi^{2} }  \int^{1}_{-1}
    dx \frac{ (1-x^{2})^{5} }{ (1 +x^{2} )^{4} } \frac{ 1}{ 2 x^{2} } \log \tilde{\Lambda} 
\end{eqnarray}

  It can be shown that the two coefficients in front of $ \gamma_{0} k_{0} $ and $ \gamma_{i} k_{i} $
  are exactly the same as expected from Lorentz invariance.
  Regularizing the integral as depicted in Fig.3,  we find that both are equal to:

\begin{equation}
    \int^{1}_{-1} dx \frac{ ( 1-x^{2} )^{5} }{ (1 +x^{2} )^{4} } \frac{1}{ 2 x^{2} } = -\frac{16}{3}
\end{equation}

 Putting back the prefactor $ -i \frac{16}{N} $,
  we get exactly the same anomalous dimension in Eqn. \ref{exp2}. This demonstrates that it is indeed
  a universal constant independent of renormalization scheme.

\section{ Calculation in Coulomb gauge }

    In this section, we will first evaluate the conventional Green function Eqn.\ref{non} in Coulomb gauge
   $ \partial_{i} a_{i} =0 $ and speculate how to calculate the gauge invariant one Eqn.\ref{inv} in this gauge.
   In FQH transitions as discussed in Ref. \cite{subir}, because the Coulomb interaction between the fermions
   breaks Lorentz invariance, Lorentz invariant gauges are not convenient,
   Coulomb gauge has to be employed to perform RG calculations.
  As in temporal gauge, we also run into IR divergence. However, in contrast to the Green function in temporal gauge,
  the Green function in Coulomb gauge does not correspond to any gauge invariant quantity, the IR divergence may not
  be regulariable. This is indeed the case as stressed in Ref.\cite{subir} in FQH transitions contexts
  and as demonstrated in the following in QED3 context.

   With the notations $ K_{\mu} = (\vec{k}, k_{0} ) $, in Coulomb gauge, the propagators are
\begin{eqnarray}
 < a_{0}(-K) a_{0} ( K ) > & = & \frac{K}{ k^{2} }   \nonumber   \\
 < a_{i}(-K) a_{j} ( K ) > & = & ( \delta_{ij} - \frac{ k_{i} k_{j} }{ k^{2} } ) \frac{1}{ K }
\end{eqnarray}

    The two fermion self-energy diagrams corresponding to the two propagators are

\vspace{-1.5cm}

\epsfig{file=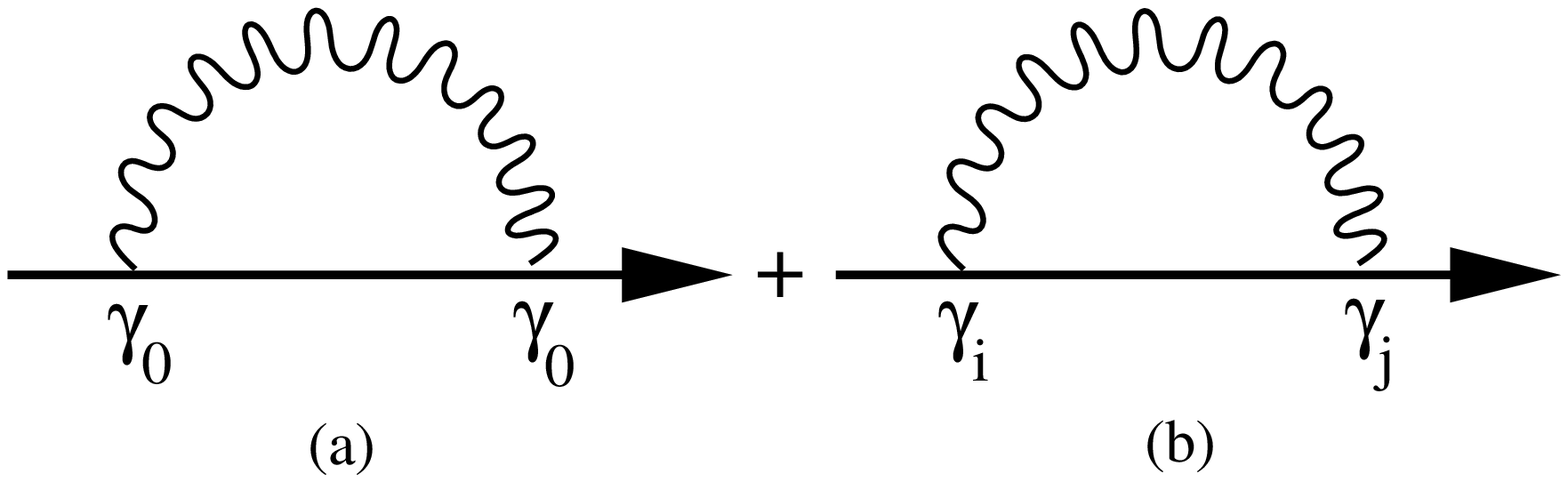,width=3.2in,height=2.5in,angle=0}

\vspace{-2.0cm}

{\footnotesize {\bf Fig 4:} The fermion self-energy diagrams in Coulomb gauge }

\vspace{0.25cm}

  The expression for Fig.4a is:
\begin{equation}
   \Sigma_{a} ( K )= -i \frac{16}{N} \int \frac{ d^{3} Q }{ (2 \pi)^{3} }
     \gamma_{0} \frac{ \gamma_{\mu} (K-Q)_{\mu} }{ (K-Q)^{2} } \gamma_{0}
    \frac{Q}{q^{2}}
\label{coul1}
\end{equation}

  The expression for Fig.4b is:
\begin{equation}
   \Sigma_{b} ( K )= -i \frac{16}{N} \int \frac{ d^{3} Q }{ (2 \pi)^{3} }
     \gamma_{i} \frac{ \gamma_{\mu} (K-Q)_{\mu} }{ (K-Q)^{2} } \gamma_{j}
    (\delta_{ij}- \frac{ q_{i} q_{j} }{ q^{2}} ) \frac{1}{Q}
\label{coul2}
\end{equation}
 
     In the following, we will try to evaluate
     $ \Sigma ( K )= \Sigma_{a} (K) + \Sigma_{b}( k) $ by two different cut-offs.

\subsection{ Cut-off in momentum and energy $ Q < \Lambda $}

 We choose the external momentum $ K $ along $ z $ axis.
 Adding the logarithmic divergences from Eqn. \ref{coul1} and \ref{coul2}, we find:
\begin{equation}
 -\frac{\gamma_{0} k_{0} } { (2 \pi)^{2} } \int^{1}_{-1} d x \frac{ (1-2 x^{2} ) x^{2} }{ 1 - x^{2} } \log \Lambda
\label{ccut1}
\end{equation}

    Again, we run into IR divergences at $ x= \pm 1 $ which are at the two end points of
  the contour integral on the real axis from $ -1 $ to $ 1 $. Unfortunately,
  from physical perscription, we are unable to 
  avoid the IR singularities at the two end points $ x=\pm 1 $ by deforming the contour.
  Therefore, we are unable to identify the anomalous exponent.

\subsection{ Cut-off in momentum space $ q < \tilde{\Lambda} $ only}
   
   In this subsection, we introduce cut-off $ \tilde{\Lambda} $ only in momentum space
  $ \vec{q} $, but integrate the frequency $ q_{0} $
  freely from $ -\infty $ to $ \infty $.

  Extracting the logarithmic UV divergences, we find:
\begin{eqnarray}
    & - &  \frac{\gamma_{0} k_{0}}{  \pi^{2}} ( \int^{1}_{-1}
    dx  \frac{ 1- 6 x^{2} + x^{4} }{ (1-x^{2}) (1 +x^{2} )^{2} }-\frac{1}{6} )
     \log \tilde{\Lambda}             \nonumber    \\
    & + & \frac{ \gamma_{i} k_{i} }{  \pi^{2} } (  \int^{1}_{-1}
    dx  \frac{ 2 x^{2} }{ (1-x^{2} ) (1 +x^{2} )^{2} } -\frac{1}{3} ) \log \tilde{\Lambda} 
\label{ccc}
\end{eqnarray}

  It can be shown that the two coefficients in front of $ \gamma_{0} k_{0} $ and $ \gamma_{i} k_{i} $
  are exactly the same as expected from Lorentz invariance. However, in contrast to temporal gauge,
  the two coefficients differ from that in Eqn.\ref{ccut1}. This causes no disturbance, because
  the Green function Eqn.\ref{non} in Coulomb gauge does not correspond to any physical quantities.
  As shown in the context of FQH transitions in Ref. \cite{subir}, this IR divergence disappears
  in physical quantities like $ \beta $ function and critical exponents. By using the same method 
  developed in Ref.\cite{subir}, it is straightforward to show that the same phenomenum happens
  here in the context of QED3 \cite{qcd}.

    In order to calculate the gauge-invariant Green function Eqn.\ref{inv} in Coulomb gauge, we could 
  extend the method developed in the following section for Lorentz covariant gauge to Coulomb gauge.
  We expect the IR divergencies in Eqns. \ref{ccut1}, \ref{ccc} to be cancelled by the inserted Dirac
  string in the gauge-invariant Green function.

\section{ Lorentz covariant calculation}

    In this section, we will calculate the gauge invariant Green function directly  in Lorentz
   covariant gauge Eqn.\ref{alpha} without
   resorting to the gauge dependent Green function. We will also compare with the result
   achieved in temporal gauge.

   The inserted Dirac string in Eqn. \ref{inv} can be written as:
\begin{equation}
   \int^{x_{2}}_{x_{1}} a_{\mu}( \xi ) d \xi_{\mu} = \int a_{\mu}(x) j^{s}_{\mu}(x) d^{d} x
\end{equation}
    where the source current is:
\begin{equation}
    j^{s}_{\mu}(x) = \int_{ {\cal C} } d \tau \frac{ d \xi_{\mu} }{ d \tau} \delta ( x_{\mu}-\xi_{\mu} (\tau) )
\end{equation}
     where $ \tau $ parameterizes the integral path $ {\cal C} $ from $ x_{1} $ to $ x_{2} $.
     
   By combining the source current with the fermion current
   $ j_{\mu} (x) = \bar{\psi}(x) \gamma_{\mu} \psi(x) $ to form the total current
   $ j^{t}_{\mu}( x ) = j_{\mu} (x) + j^{s}_{\mu}(x) $, we can write
   the gauge invariant Schwinger Green function Eqn.\ref{inv} as:
\begin{eqnarray}
  G^{inv}(x_{1}, x_{2} ) & = & \frac{1}{ Z } \int {\cal D} \psi {\cal D} \bar{\psi} \psi(x_{1}) \bar{\psi}(x_{2})
      e^{- \int d^{d} x \bar{\psi} \gamma_{\mu} \partial_{\mu} \psi}     
                                     \nonumber   \\
     & \cdot & \int {\cal D} a_{\mu} e^{-\int d^{d} x ( S [ a_{\mu} ] - i e j^{t}_{\mu}(x) a_{\mu}(x) )}
\end{eqnarray}
     Where $ Z $ is the partition function of QED3:
\begin{equation}
  Z =  \int {\cal D} \psi {\cal D} \bar{\psi} {\cal D} a_{\mu}
     e^{- \int d^{d} x ( \bar{\psi} \gamma_{\mu} ( \partial_{\mu} -i e a_{\mu} ) \psi + S[ a_{\mu} ] ) }
\end{equation}

     Because $ G^{inv}(x,y) $ is a gauge-invariant quantity, we can integrate out the gauge field $ a_{\mu} $
  in any fixed gauge without affecting the final result. For simplicity, we choose
   Landau gauge $ \alpha =0 $ in Eqn.\ref{alpha}. Extension to any gauge fixing parameter is straightforward
   and results will be given at the end of the section. Integrating out $ a_{\mu} $
  in Landau gauge  leads to
\begin{eqnarray}
  G^{inv}(x_{1}, x_{2}) &  = &  \frac{1}{ Z } \int {\cal D} \psi {\cal D} \bar{\psi} \psi(x_{1}) \bar{\psi}(x_{2})
      e^{- \int d^{d} x \bar{\psi} \gamma_{\mu} \partial_{\mu} \psi }  
                                                             \nonumber  \\
     & \cdot & e^{- \frac{ e^{2} }{ 2 } \int d x d x^{\prime}
           j^{t}_{\mu}(x) D_{\mu \nu}(x-x^{\prime}) j^{t}_{\nu}( x^{\prime} ) }
\end{eqnarray}

   After we expand the total current in terms of the fermion current and the source current, the last
   integrand becomes
\begin{eqnarray}
        &  &  j_{\mu}(x) D_{\mu \nu}(x-x^{\prime}) j_{\nu}( x^{\prime} )
                       \nonumber   \\
         & +  &   j^{s}_{\mu}(x) D_{\mu \nu}(x-x^{\prime}) j^{s}_{\nu}( x^{\prime} )
                       \nonumber   \\
         & +  &   2 j_{\mu}(x) D_{\mu \nu}(x-x^{\prime}) j^{s}_{\nu}( x^{\prime} )  
\label{three}
\end{eqnarray}
    
    The first term in Eqn.\ref{three}
    is just the conventional long-range four-fermion interaction mediated by the gauge field,
    it leads to the anomalous exponent in  Landau gauge given in section II:
\begin{equation}
   \eta_{1}= \frac{16}{N} \frac{1}{ 3 \pi^{2} }
\end{equation}

       The second term is equal to:
\begin{eqnarray}
     & - & \frac{16}{ N (2 \pi)^{3} }
     \int \frac{d^{3} k }{ k^{3}} \frac{ k^{2} (x_{2}-x_{1})^{2}- (k \cdot (x_{2}-x_{1}) )^{2} }
      { ( k \cdot (x_{2}-x_{1}) )^{2} }    \nonumber   \\ 
       & \cdot & (1- \cos k \cdot (x_{2}-x_{1}) ) 
                       \nonumber  \\
    & = & - \frac{16}{ (2 \pi)^{2} N} \int^{\Lambda}_{0} \frac{ d k} { k } 
    \int^{1}_{-1} d x \frac{ 1-x^{2} }{ x^{2} } ( 1- \cos k r x)
         \nonumber    \\
    & = & - \frac{ 8 }{ \pi^{2} N }  ( -2 \gamma -2 \log \Lambda r + \cos \Lambda r + \frac{\pi}{2} \Lambda r )
\end{eqnarray}
     Where $ r= | x_{2}-x_{1} | $ and $ \gamma $ is the Euler constant.

      The linear divergence is a arti-fact of the momentum cut-off and should be ignored in Lorentz invariant
   regularization. The $ \cos \Lambda r $ is a fast oscillating factor and averages to zero \cite{free}. The logarithmic
  factor leads to the anomalous exponent:
\begin{equation}
   \eta_{2}= \frac{16}{N} \frac{1}{  \pi^{2} }
\end{equation}
       which has the {\em same} sign as , but three times larger than $ \eta_{1} $.

       We can write the third term as
\begin{equation}
    -e^{2}  \int d x \bar{\psi}(x) \gamma_{\mu} \psi(x) \int^{x_{2}}_{x_{1}} d x^{\prime}_{\nu} D_{\mu \nu}(x-x^{\prime})
\label{third}
\end{equation}

     Eqn. \ref{third} is essentially quadratic in the fermions. Combining it with the free fermion action leads to:
\begin{equation}
   {\cal S}= \int d x \bar{\psi}(x) ( \gamma_{\mu} \partial_{\mu} +
    e^{2} \gamma_{\mu} \int^{x_{2}}_{x_{1}} d x^{\prime}_{\nu} D_{\mu \nu}(x-x^{\prime}) ) \psi(x)
\end{equation}

    In principal, the propagator of fermion $ < \psi(x_{1}) \bar{\psi}(x_{2}) > $ can be calculated by inverting
  the quadratic form in the above equation, but it is not easy to carry out in practice. Instead
  we can construct perturbative expansion in {\em real space} by the following Feymann diagrams in Fig.5

\vspace{-1.5cm}

\epsfig{file=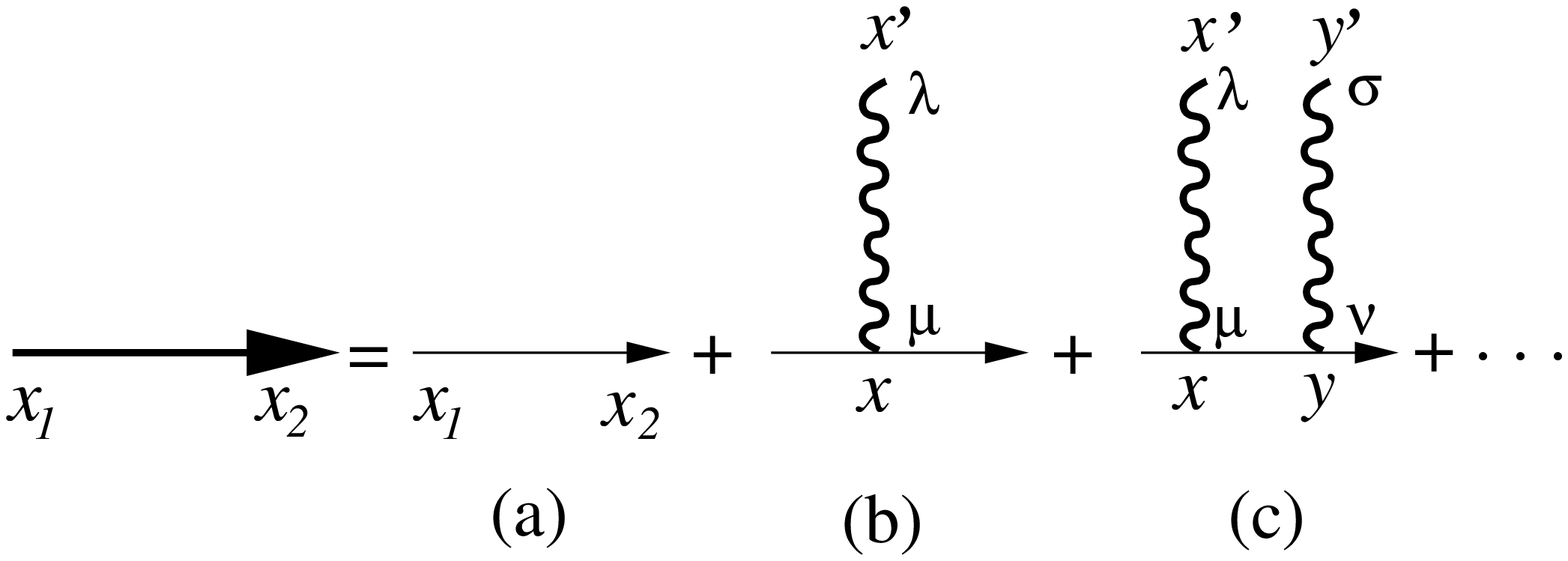,width=3.2in,height=2.5in,angle=0}

\vspace{-2.0cm}

{\footnotesize {\bf Fig 5:} The perturbative expansion series of Eq.\ref{third} }

\vspace{0.25cm}

  The corresponding expression is:
\begin{eqnarray}
  & & G(x_{1},x_{2})= G_{0}(x_{1},x_{2})   \nonumber   \\
  &- & e^{2} \int dx G_{0}(x_{1},x)
  \int^{x_{2}}_{x_{1}} d x^{\prime}_{\lambda} \gamma_{\mu} D_{\mu \lambda}(x-x^{\prime}) G_{0}(x, x_{2})
                \nonumber    \\
  & + & e^{4} \int dx G_{0}(x_{1},x)
  \int^{x_{2}}_{x_{1}} d x^{\prime}_{\lambda} \gamma_{\mu} D_{\mu \lambda}(x-x^{\prime})
                 \nonumber \\
  & \cdot  & \int dy G_{0}(x,y)
  \int^{x_{2}}_{x_{1}} d y^{\prime}_{\sigma} \gamma_{\nu} D_{\nu \sigma}(y-y^{\prime})
  G_{0}(y, x_{2}) + \cdots
\end{eqnarray}

  Being just quadratic, there is no loops in the above Feymann series.
  But there may still be potential divergences.
    
      The explicit expression for Fig.5b is
\begin{eqnarray}
    F(x) & = & - e^{2} \int \frac{ d^{3} q_{1}}{ (2 \pi )^{3} } \frac{ d^{3} q_{2}}{ (2 \pi )^{3} }
        \frac{ e^{-i q_{1} x}- e^{-i q_{2} x} }{ -i ( q_{1}- q_{2} ) \cdot x }   \nonumber   \\
       & \cdot &  G_{0}( q_{1} ) \gamma_{\mu} x_{\nu} D_{\mu \nu}(q_{1}-q_{2})G_{0}(q_{2})
\label{fx}
\end{eqnarray}
     where $ x_{2}-x_{1} = x $.

  Suppressing the prefactor $ \frac{16}{N} $, we can write Eqn.\ref{fx} as the sum of two parts
  $ F(x)= F_{1}(x) + F_{2}(x) $ with
\begin{eqnarray}
    F_{1}(x) & = &  2 i \int \frac{ d^{3} k}{ (2 \pi )^{3} }
        \frac{ e^{i k x} }{ k^{2} } \int  \frac{ d^{3} q}{ (2 \pi )^{3} }    
    (  \frac{ \gamma_{\mu}  k_{\mu} } { q ( q-k)^{2} }     \nonumber  \\
   & + & \frac{ -2 k \cdot x \gamma_{\mu} k_{\mu} + k^{2} \gamma_{\mu} x_{\mu} }
            { q \cdot x q ( q-k)^{2} }   \nonumber  \\ 
    & + & \frac{ k \cdot x \gamma_{\mu} q_{\mu} - k \cdot q \gamma_{\mu} x_{\mu} }
            { q \cdot x q ( q-k)^{2} }  )   
\label{fx1}
\end{eqnarray}
     and
\begin{equation}
    F_{2}(x)  =  - 2 i \int \frac{ d^{3} k}{ (2 \pi )^{3} }
        \frac{ e^{i k x} }{ k^{2} } \int \frac{ d^{3} q}{ (2 \pi )^{3} }
       \frac{ k^{2} \gamma_{\mu} q_{\mu}+ q^{2} \gamma_{\mu} k_{\mu} }{ (k+q)^{3} q^{2} } 
\label{fx2}
\end{equation}

  The logarithmic divergence in the first term in Eqn.\ref{fx1}
  cancels exactly that of Eqn.\ref{fx2}.
  The second term in Eqn.\ref{fx1} is convergent.
  So the only possible divergence may come from the third term in Eqn.\ref{fx1}:
\begin{equation}
    \int  \frac{ d^{3} q}{ (2 \pi )^{3} } 
    \frac{ k \cdot x \gamma_{\mu} q_{\mu} - k \cdot q \gamma_{\mu} x_{\mu} }
    { q \cdot x q ( q-k)^{2} }
\label{x}
\end{equation}

   It is easy to find the UV divergences in the first term and second term in
   Eqn.\ref{x} exactly cancel each other as follows:
\begin{equation}
    - \frac{ x \cdot k \gamma_{\mu} x_{\mu} }{ 2 \pi^{2} x^{2} } \log \Lambda
    + \frac{ x \cdot k \gamma_{\mu} x_{\mu} }{ 2 \pi^{2} x^{2} } \log \Lambda =0
\end{equation}  

    We conclude that there is no extra divergence from $ F(x) $ in Eqn.\ref{fx}. To order $ 1/N $, we have
\begin{equation}
   \eta_{3}= 0
\label{app}
\end{equation}

      In all, the final anomalous exponent is
\begin{equation}
   \eta= \eta_{1} + \eta_{2} + \eta_{3}= \frac{64 }{ 3 \pi^{2} N } 
\label{final}
\end{equation}

      This is exactly the same as that calculated in the temporal gauge.

   It is straightforward to repeat the above calculation by using different gauge fixing parameter $ \alpha $
   in Eqn.\ref{alpha}, then $ \eta_{1}, \eta_{2}, \eta_{3} $ all depend on $ \alpha $ separately in
   the following way: $ \eta_{1}= \frac{16}{N} \frac{1}{ 3 \pi^{2} }- \frac{ 8 \alpha }{ N \pi^{2} },
   \eta_{2}= \frac{16}{N} \frac{1}{  \pi^{2} }- \frac{ 8 \alpha }{ N \pi^{2} },
   \eta_{3}=  \frac{ 16 \alpha }{ N \pi^{2} } $. It is easy to see that $\alpha $ disappears 
   in the final answer $ \eta=\eta_{1} + \eta_{2} + \eta_{3} = \frac{64}{ 3 \pi^{2} N} $.

   There are many $ (1/N)^{2} $ corrections.
   $ \eta_{1} $ has the standard correction, $ \eta_{2} $ has the correction from the propagator in Eqn.\ref{alpha},
   $ \eta_{3} $ has the correction from Fig.5c  and the new anomalous exponent $ \eta_{4} $ has the correction
   from Fig.6 which is the lowest order mixing diagram between the first term and the third term in Eqn. \ref{three}.
\vspace{-1.1cm}

\epsfig{file=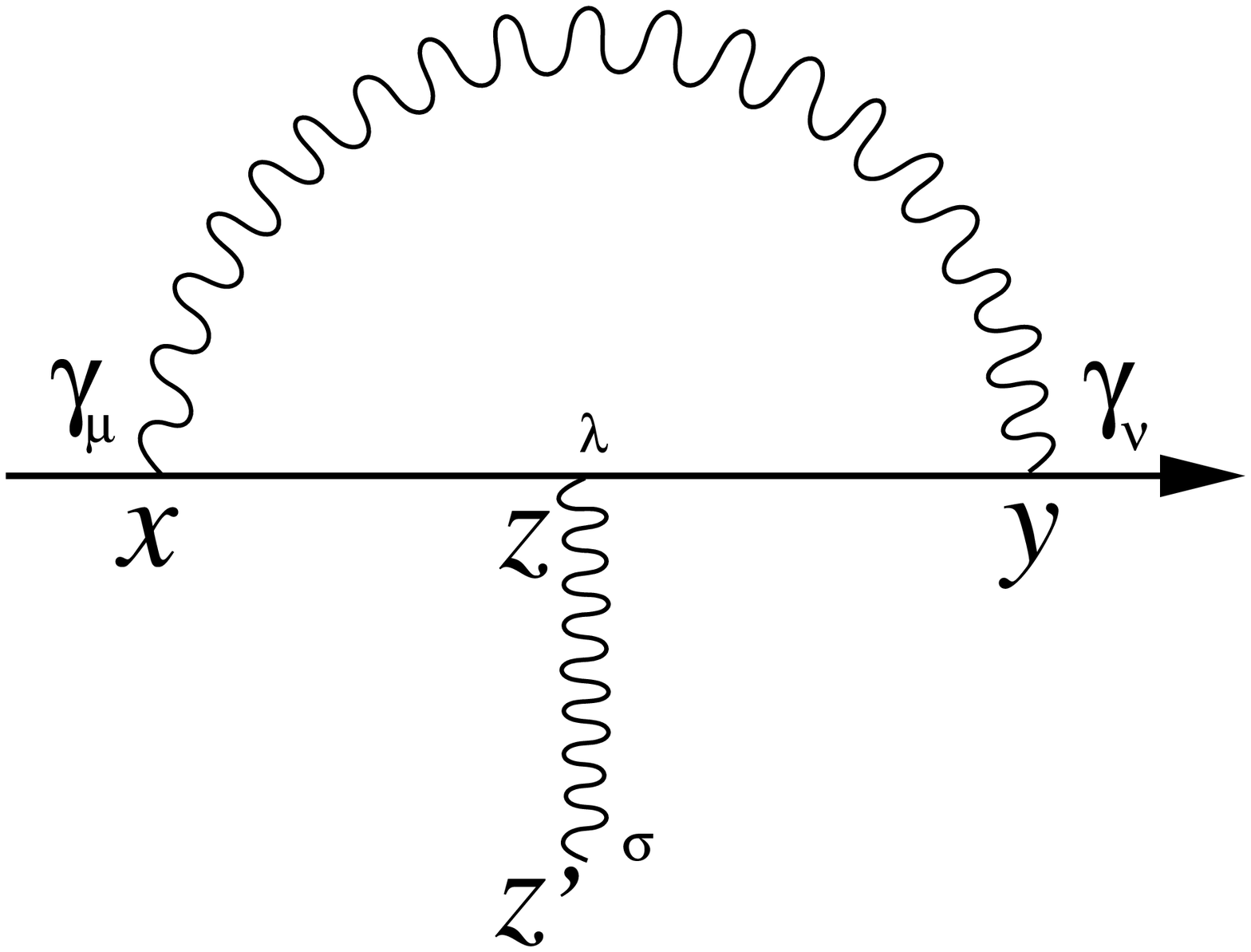,width=3.2in,height=2.5in,angle=0}

\vspace{-1.6cm}

{\footnotesize {\bf Fig 6:} The lowest order mixing between the first and the third term }

\vspace{0.25cm}

    Although evaluating all these $ (1/N)^{2} $ corrections is beyond the scope of this paper, we have
  established firmly our results Eqn.\ref{final} to order $ 1/N $.
\section{conclusions}

  In this paper, we calculated the gauge-invariant fermion Green function in $ 2+1 $ dimensional QED
  by different methods.
   The calculations in temporal gauge with different cut-offs in section III and Lorentz covariant
  calculation with different gauge fixing parameters $ \alpha $ in section V all lead to the same answer.
  These facts strongly suggest Eqn. \ref{exp2} is
  the correct exponent in the leading order of $ 1/ N $.
  These methods have been applied to many very different physical systems \cite{bose}.

   Finally, we would like to comment on the possible application of the results of this paper on high temperature
  superconductors.  As explicitly demonstrated in Refs. \cite{thermal,quantum},
  in the vortex plasma regime in the neighborhood of the finite temperature
  phase transition boundary, the relevant gauge-invariant fermion Green function is that of fermions moving in
  classical random magnetic field generated by the vortex plasma. Obviously, this static gauge-invariant
  Green function is expected to show rather different behaviors than the dynamic one calculated in this paper.
  For technical and simplicity reason, the dynamic gauge-invariant one was used in Ref.\cite{thermal} to present the
  Energy Distribution Curve (EDC) and Momentum Distribution Curve (MDC) of ARPES data. We used the anomalous
  exponent calculated by RW which differs from our result Eqn. \ref{exp2} both in sign and in magnitude.
  However, if correct anomalous exponent is used in the calculation in Ref.\cite{thermal} is irrelevant anyway, because 
  as cautioned at the end of Ref.\cite{thermal}, the static gauge-invariant
  Green function should be used  to compare with experimental ARPES data \cite{bose}.

 This work was supported by the Pennsylvania state University.
 I thank Ahsok Sen for numerous helpful discussions.
 I also thank R. Jackiw, J. K. Jain and X. G. Wen for helpful discussions.

  Note added in proof. After we achieved the results of this paper, we got to know the preprint 
  cond-mat/0112202 by Khveshchenko. He used very different methods and got the same anomalous exponent as mine
  Eqn.\ref{exp2} upto a factor of 2. I suspect that the factor of 2 discrepancy may be due to he used
  4 component spinors in contrast to the two component spinors used in this paper. Four component spinor representation
  has the advantage to investigate dynamical parity-conserving mass generation as discussed in Refs.\cite{qed,brad}.

\end{multicols}

\begin{references}
\bibitem{geo} Because $ x=-\cos \theta $ is dimensionless, it may be more suitable to call
    the divergence geometric divergence.
\bibitem{free} For  free field $ G_{0}(k) =\frac{ i \gamma_{\mu} k_{\mu} }{ k^{2} } $. In real space,
   it corresponds to $ \frac{ \gamma_{\mu} x_{\mu}} { 2 \pi^{2} x^{3} } ( \sin \Lambda r -\frac{\pi}{2} ) $.
   In $ \Lambda \rightarrow \infty $, the fast oscillating piece averages zero and can be dropped.
\bibitem{qcd} That IR divergencies should be cancelled in any physical quantities was summarized 
      in Quantum Chromodynamics ( QCD ) as two theorems: Kinoshita-Poggio-Quinn theorem and Kinoshita-Lee-nauenberg theorem.
      For non-relativistic quantum field theories, it was discussed in the context of FQH effect
      in Ref.7 and Ref. 8 and in the context of superconductor to insultaor transition in Ref 9
\end{references}
\end{document}